\documentclass[aps, prl, twocolumn]{revtex4-2}
\usepackage{amssymb, amsfonts, amsmath, amsthm, mathtools, mathrsfs}
\usepackage{txfonts, pxfonts, upgreek}
\usepackage[usenames, dvipsnames]{xcolor}
\usepackage{braket}
\usepackage[colorlinks, linkcolor=blue, anchorcolor=red, citecolor=blue]{hyperref}
\usepackage{natbib}
\usepackage{array}
\usepackage{bigstrut}
\usepackage{inputenc}
\usepackage[T1]{fontenc}
\usepackage{lmodern}
\usepackage[english]{babel}

\begin{document}

\title{Tensor-Network Approach to Work Statistics for 1D Quantum Lattice Systems}

\author{Jiayin Gu\textsuperscript{1}}\email{gujiayin@pku.edu.cn}
\author{Fan Zhang\textsuperscript{1}}
\author{H. T. Quan\textsuperscript{1,2,3}}\email{htquan@pku.edu.cn}

\affiliation{\textsuperscript{\rm 1}School of Physics, Peking University, Beijing 100871, China;}
\affiliation{\textsuperscript{\rm 2}Collaborative Innovation Center of Quantum Matter, Beijing 100871, China;}
\affiliation{\textsuperscript{\rm 3}Frontiers Science Center for Nano-Optoelectronics, Peking University, Beijing 100871, China}

\begin{abstract}
We introduce a numerical approach to calculate the statistics of work done on 1D quantum lattice systems initially prepared in thermal equilibrium states. This approach is based on two tensor-network techniques: Time Evolving Block Decimation (TEBD) and Minimally Entangled Typical Thermal States (METTS). The former is an efficient algorithm used to simulate the dynamics of 1D quantum lattice systems, while the latter a finite-temperature algorithm for generating a set of typical states representing the Gibbs canonical ensemble. As an illustrative example, we apply this approach to the Ising chain in mixed transverse and longitudinal fields. Under an arbitrary protocol, the moment generating function of the work can be obtained, from which the work moments are numerically calculated and the quantum Jarzynski equality can be tested. Moreover, the numerical approach is also adapted to test the universal quantum work relation involving a functional of an arbitrary observable.
\end{abstract}

\maketitle

\par Microreversibility, a fundamental symmetry of Nature, dictates various nonequilibrium relations, which are nowadays collectively known as fluctuation theorem~\cite{Bochkov_SovPhysJETP_1977, Evans_PhysRevLett_1993, Gallavotti_PhysRevLett_1995, Jarzynski_PhysRevLett_1997, Kurchan_JPhysA_1998, Crooks_PhysRevE_1999, Lebowitz_JStatPhys_1999, Maes_JStatPhys_1999, Seifert_EurPhysJB_2008}. Among these relations, the Jarzynski equality attracts considerable interest. It is a parameter-free, model-independent relation, and allows to express the free energy difference between two equilibrium states by a nonlinear average over the required work to drive the system in a nonequilibrium process from one state to another. Over the last decades, extensive efforts were devoted to prove and experimentally test the Jarzynski equality or closely related Crooks fluctuation theorem in various systems~\cite{Tasaki_arXiv_2000, Kurchan_arXiv_2000, DeRoeck_PhysRevE_2004, Talkner_PhysRevE_2007, Huber_PhysRevLett_2008, Jarzynski_EurPhysJB_2008, Jarzynski_AnnuRevCondensMatterPhys_2011, Esposito_RevModPhys_2009, Campisi_RevModPhys_2011, Dorner_PhysRevLett_2013, Mazzola_PhysRevLett_2013, Klages_2013, Batalhao_PhysRevLett_2014, Hummer_ProcNatlAcadSciUSA_2010, Liphardt_Science_2002, Collin_Nature_2005, An_NatPhys_2015}. However, what's more informative is the detailed probability distribution of work under an arbitrary protocol (instead of a sudden quench)~\cite{Arrais_PhysRevE_2018, Arrais_PhysRevE_2019}. Since it encodes essential information about not only the equilibrium properties, but also the nonequilibrium driving processes~\cite{Heyl_PhysRevLett_2013, Silva_PhysRevLett_2008, Dora_PhysRevB_2012, Russomanno_JStatMech_2015, Jarzynski_PhysRevX_2015, Zhu_PhysRevE_2016, Wang_PhysRevE_2017, Goold_incollection_2018, Beyer_PhysRevRes_2020}. For quantum many-body systems, the reality is that it is formidably challenging to calculate the work statistics under an arbitrary driving protocol. Previous studies mainly focus on analytical methods and are restricted to few exactly solvable models, and are studied case by case. Examples include harmonic oscillators~\cite{Funo_PhysRevLett_2018, Deffner_PhysRevE_2008, Talkner_PhysRevE_2008b, Jaramillo_PhysRevE_2017, Yadalam_PhysRevA_2019, Myers_PhysRevE_2020}, piston systems~\cite{Qiu_PhysRevE_2020, Quan_PhysRevE_2012, Gong_PhysRevE_2014}, 1D diatomic Toda lattice~\cite{Zhu_PhysRevE_2018}, 1D quantum gases~\cite{Gong_PhysRevE_2014, Zheng_PhysRevE_2015, Wang_PhysRevE_2018, Atas_PhysRevA_2020}, quantum fields~\cite{Ortega_PhysRevLett_2019, Bartolotta_PhysRevX_2018}, and quantum systems of quadratic Hamiltonians~\cite{Smacchia_PhysRevE_2013, Fei_PhysRevA_2019, Fei_PhysRevRes_2019, Scandi_PhysRevRes_2020, Varizi_PhysRevRes_2020}. Quantum Feynman-Kac equation~\cite{Liu_PhysRevE_2012} and phase-space formulation~\cite{Fei_PhyRevE_2018, Qian_PhysRevE_2019, Brodier_JPhysA_2020} have been proposed, but are practically constrained to single-particle systems and difficult to extend to many-body systems. Nonequilibrium Green's function approach~\cite{Fei_PhysRevLett_2020b}, group-theoretical approach~\cite{Fei_PhysRevA_2019, Fei_PhysRevRes_2019}, and path-integral approach~\cite{Funo_PhysRevLett_2018, Dong_PhysRevB_2019, Aron_SciPostPhys_2018, Qiu_PhysRevE_2020, Yeo_PhysRevE_2019} have been proposed, but only applicable to perturbative driving protocols and/or quadratic Hamiltonians. Despite of these efforts, a systematic methods for calculating the work distribution of a quantum many-body system under an arbitrary protocol is still lacking, thus demanding to develop numerical ones. The tensor-network approach~\cite{Montangero_2018, Orus_NatRevPhys_2019} is such an ideal candidate, which drastically decrease the computation complexity associated with the exponentially-large Hilbert space intrinsic to  quantum many-body systems. Although originally developed in the context of condensed matter physics, tensor-network approach is increasingly being applied to tackle problems in other fields of research. In quantum thermodynamics, for example, tensor-network approach has recently been used to simulate strongly interacting quantum thermal machines~\cite{Brenes_PhysRevX_2020} and to study the heat transfer in non-Markovian open quantum systems~\cite{Popovic_PRXQuantum_2021}.

\par Based on two tensor-network techniques, Time Evolving Block Decimation (TEBD)~\cite{Paeckel_AnnPhys_2019} and Minimally Entangled Typical Thermal States (METTS)~\cite{White_PhysRevLett_2009}, we introduce in this Letter an efficient numerical approach to calculate the work statistics for 1D quantum lattice systems in nonequilibrium processes. The quantum Ising chain in the presence of transverse and longitudinal fields is chosen to benchmark this approach.

\par \textit{Two-point measurement scheme}.-- In the nanoscopic world, extension of classical Jarzynski equality to quantum systems can be realized  by a proper definition of work introduced in the year 2000~\cite{Kurchan_arXiv_2000, Tasaki_arXiv_2000}. In this scheme, a measurement of energy is initially performed on the system in thermal equilibrium state. In the sequel, the system evolves under an external driving force before another measurement of energy at the final time $\tau$. The fluctuating work is defined as the energy difference between the two eigenenergies, $W_{m,n}=E_m^{\tau}-E_n^0$. The joint probability of observing such measured energies is given by ${\cal P}(n, m)={\cal P}_n|\braket{m(\tau)|U|n(0)}|^2$, where $\ket{n(t)}$ is the $n$-th instantaneous energy eigenstate of the system at time $t$, ${\cal P}_n=\braket{n(0)|\rho|n(0)}$ the initial probability of $\ket{n(0)}$, $\rho={\rm e}^{-\beta H(0)}/Z$ the initial Gibbs canonical density matrix of the system at the inverse temperature $\beta\equiv 1/(k_{\rm B}T)$, and $k_{\rm B}$ the Boltzmann's constant. Besides, $U$ denotes the unitary operator ruling the time evolution of the system, which is expressed in terms of the time-dependent Hamiltonian $H(t)$ and the time-ordering operator ${\cal T}$,
\begin{align}
U={\cal T}\exp\left[-\frac{\rm i}{\hbar}\int_0^{\tau}H(t)\,{\rm d}t\right] \text{.}
\end{align}
The work distribution is therefore given by ${\cal P}(W)=\sum_{m,n}\delta(W-W_{m,n}){\cal P}(m, n)$. We define the moment generating function of the work distribution,
\begin{align}
G(s)=\int {\cal P}(W)\,{\rm e}^{sW}{\rm d}W \text{,} \label{eq_G_s_defined}
\end{align}
then the moment generating function can be expressed as
\begin{align}
G(s)={\rm Tr}\left[U^{\dagger}{\rm e}^{sH(\tau)} U{\rm e}^{-sH(0)}\rho\right] \text{.} \label{eq_G_s}
\end{align}
The moments of work can be obtained by taking successive derivatives,
\begin{align}
\langle W^n\rangle=\left.\frac{{\rm d}^nG(s)}{{\rm d}s^n}\right\vert_{s=0} \text{.} \label{eq_moments}
\end{align}
The moment generating function~(\ref{eq_G_s}) is the quantity that we will numerically calculate with the tensor-network approach.

\par
\begin{figure}
\begin{minipage}[t]{0.8\hsize}
\resizebox{1.0\hsize}{!}{\includegraphics{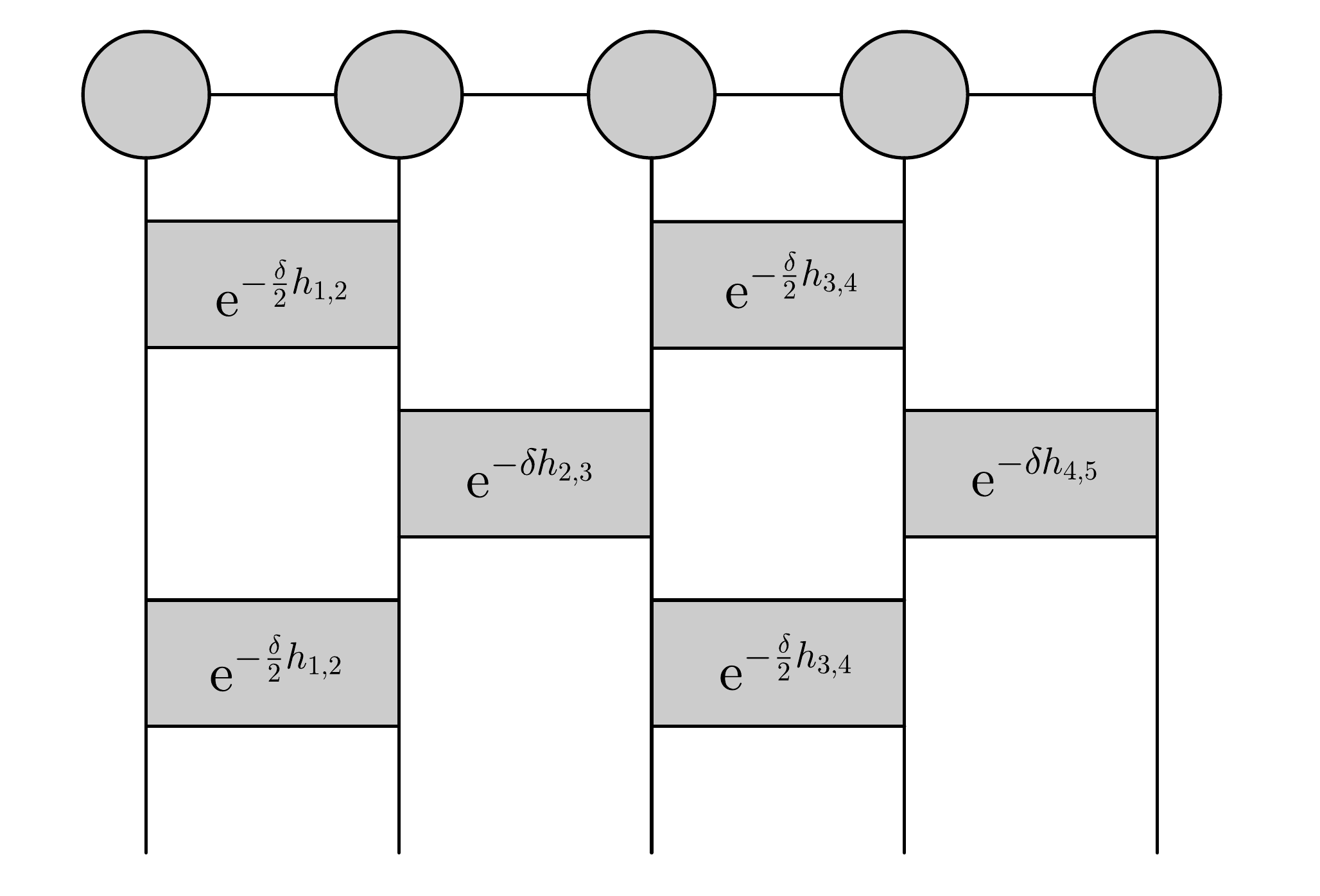}}
\end{minipage}
\caption{Diagrammatic representation of the TEBD algorithm for a quantum lattice system of $5$ sites with nearest-neighbor interactions. The full Hamiltonian is splitted into two parts, $H=H_{\rm odd}+H_{\rm even}$ with $H_{\rm odd}=h_{1,2}+h_{3,4}$ and $H_{\rm even}=h_{2,3}+h_{4,5}$. The odd and even numbered two-site local evolution operators are alternatively applied to the wave function represented by a matrix product state (MPS)~\cite{PerezGarcia_arXiv_2007}.}
\label{fig_TEBD}
\end{figure}

 \textit{Time Evolving Block Decimation}.-- TEBD is an algorithm that relies on the Trotter-Suzuki decomposition \cite{Suzuki_CommunMathPhys_1976} and subsequent approximation of the exact evolution operator $U^{\rm exact}(\delta)$ for the small quantity $\delta$. The full Hamiltonian $H$ can be splitted into $N_H$ parts, $H=\sum_{\alpha=1}^{N_H} H_{\alpha}$, where each part $H_{\alpha}$ is a sum, $H_{\alpha}=\sum_{k=1}^{N_{\alpha}}h_{\alpha}^k$, such that $h_{\alpha}^k$ can be diagonalized efficiently and are mutually commuting, $[h_{\alpha}^k,\,h_{\alpha}^l]=0$. The exact evolution operator can be decomposed to any order. Here, we give the second-order one,
\begin{align}
U^{\rm exact}(\delta) & ={\rm e}^{-\delta\sum_{\alpha=1}^{N_H}H_{\alpha}} \nonumber \\
& \approx\prod_{\alpha=1}^{N_H}{\rm e}^{-\frac{\delta}{2}H_{\alpha}}\prod_{\alpha=N_H}^1{\rm e}^{-\frac{\delta}{2}H_{\alpha}}+O(\delta^3) \text{,}
\end{align}
as it is commonly used. Figure~\ref{fig_TEBD} depicts the TEBD algorithm with diagrammatic notations.

\par
\begin{figure}
\begin{minipage}[t]{0.8\hsize}
\resizebox{1.0\hsize}{!}{\includegraphics{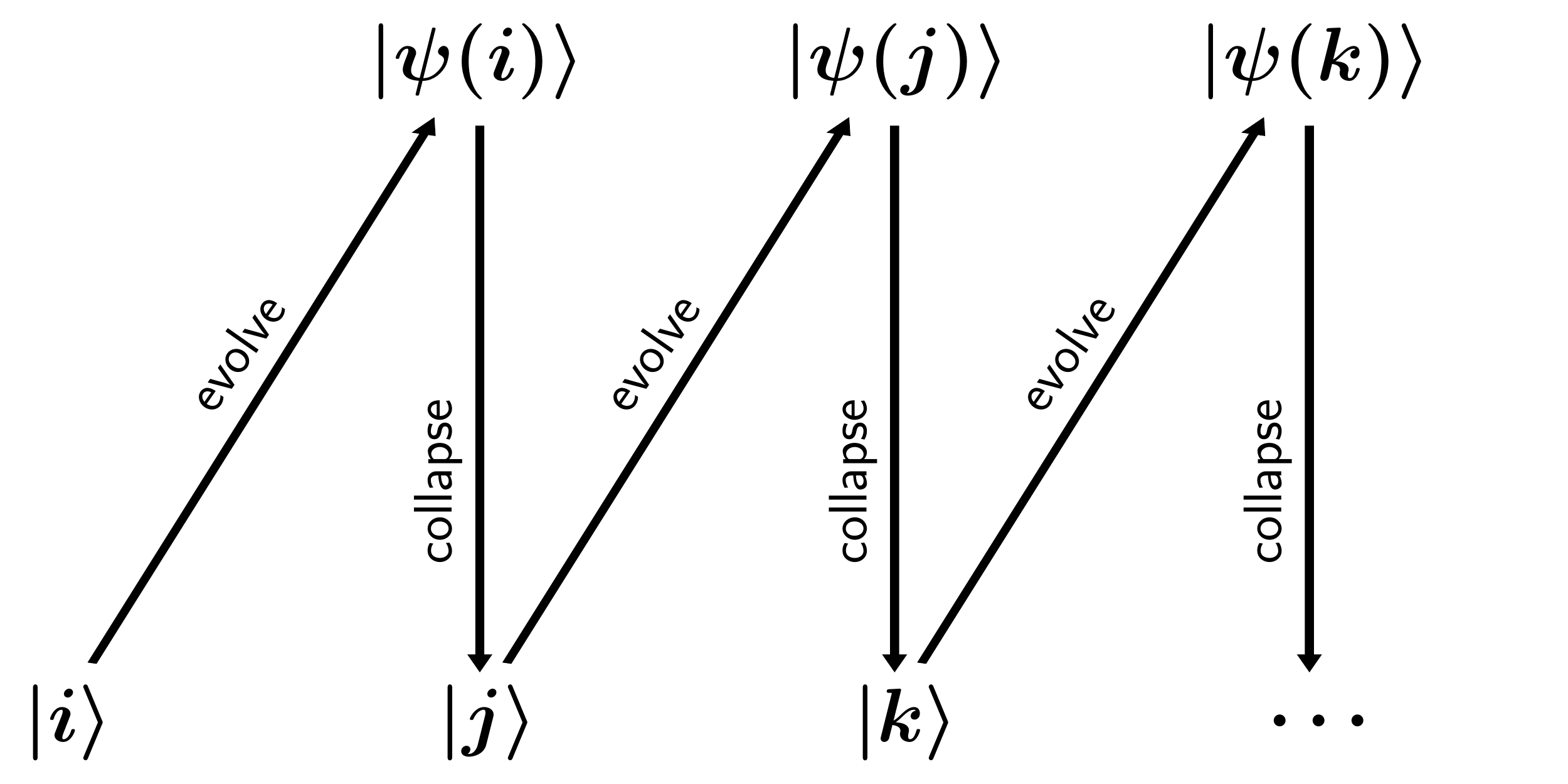}}
\end{minipage}
\caption{Schematic representation of the METTS algorithm whose procedure consists of the following steps: (i) choose a cps $\ket{i}$; (ii) evolve it to a metts $\ket{\psi(i)}$ in imaginary time according to Eq.~(\ref{eq_METTS}) and calculate quantities of interest; (iii) collapse $\ket{\psi(i)}$ into a new cps $\ket{j}$ with the probability ${\cal P}(i\to j)=|\braket{j|\psi(i)}|^2$ and then return to the step (ii).}
\label{fig_METTS}
\end{figure}

 \textit{Minimally Entangled Typical Thermal States}.-- METTS is a finite-temperature algorithm for generating a set of typical states representing the Gibbs canonical ensemble. For a quantum lattice system, starting from a product state $\ket{i}$, we can generate a typical state called metts $\ket{\psi(i)}$ with evolution in imaginary time,
\begin{align}
\ket{\psi(i)}=\frac{1}{\sqrt{{\cal P}(i)}}{\rm e}^{-\beta H/2}\ket{i} \text{,} \label{eq_METTS}
\end{align}
where ${\cal P}(i)=\braket{i|{\rm e}^{-\beta H}|i}$. Here, the evolution is realized using TEBD. A set of metts satisfy the typicality condition,
\begin{align}
\rho=\frac{{\rm e}^{-\beta H}}{Z}=\sum_i\frac{{\cal P}(i)}{Z}\ket{\psi(i)}\bra{\psi(i)} \text{,} \label{eq_rho}
\end{align}
where $Z$ denotes the partition function, and ${\cal P}(i)/Z$ is therefore the weight of $\ket{\psi(i)}$.  A thermal measurement of an arbitrary static observable $O$ can be calculated as
\begin{align}
\langle O\rangle=\frac{1}{Z}\sum_i{\cal P}(i)\braket{\psi(i)|O|\psi(i)} \text{,}
\end{align}
To sample the metts ensemble randomly according to the probability distribution ${\cal P}(i)/Z$, a Markov chain of the product state is constructed by first obtaining a metts $\ket{\psi(i)}$ from a product state $\ket{i}$, second collapsing the metts $\ket{\psi(i)}$ into a a new product state $\ket{j}$ with the probability ${\cal P}(i\to j)=|\braket{j|\psi(i)}|^2$, and then repeating this procedure. $\ket{i}$ is henceforth referred to as a collapsed product state (cps). Consider the ensemble of all cps $\ket{i}$ initially distributed with probability ${\cal P}(i)/Z$, it can be checked that the detailed balance condition is satisfied,
\begin{align}
 \frac{{\cal P}(i)}{Z}{\cal P}(i\to j)= \frac{{\cal P}(j)}{Z}{\cal P}(j\to i) \text{,}
\end{align}
guaranteeing the stability of the Markov chain. See Figure~\ref{fig_METTS} for a brief illustration of the METTS algorithm. Detailed accounts can be found in Ref.~\cite{Stoudenmire_NewJPhys_2010}.

\par \textit{Illustrative example}.-- We now consider a spin one-half quantum Ising chain of $L$ sites with nearest-neighbor interactions in the presence of transverse and longitudinal fields. The Hamiltonian is given by
\begin{align}
H=-J\sum_{\langle j,j+1\rangle}S_j^zS_{j+1}^z-h_x\sum_jS_j^x-h_z\sum_jS_j^z \text{,} \label{eq_Hamiltonian}
\end{align}
where $S_j^z$ and $S_j^x$ denote the operators at the $j$-th site defined in terms of Pauli matrices,
\begin{align}
& S_j^x=\frac{\sigma^x}{2}=\frac{1}{2}\begin{pmatrix}
0 & 1 \\
1 & 0
\end{pmatrix} \text{,}
& S_j^z=\frac{\sigma^z}{2}=\frac{1}{2}\begin{pmatrix}
1 & 0 \\
0 & -1
\end{pmatrix} \text{,}
\end{align}
and $J$ is the exchange constant. $h_x$ (respectively, $h_z$) is the magnetic field in $x$ (respectively, $z$) direction controlled externally in time according to a prescribed protocol, thus making the Hamiltonian generally time-dependent. Different from the transverse-field Ising chain, the Ising chain in mixed fields cannot be solved exactly. We have to resort to numerical approach for detailed investigation.

\par In order to fit into the TEBD algorithm, the full Hamiltonian is splitted into odd and even parts, $H=H_{\rm odd}+H_{\rm even}$, with $H_{\rm odd}$ and $H_{\rm even}$ each being the sum of mutually commuting local two-site operators,
\begin{align}
H_{\rm odd}=\sum_{j\in {\rm odd\,set}} h_{j,j+1} \text{,}\hspace{0.5cm} H_{\rm even}=\sum_{j\in {\rm even\,set}} h_{j,j+1} \text{,}
\end{align}
as illustrated in Figure~\ref{fig_TEBD}. Each local two-site operator $h_{j,j+1}$ is constructed as follows,
\begin{align}
h_{j,j+1}= & -JS_j^z\otimes S_{j+1}^z \nonumber \\
& -\frac{(1+\delta_{j,1})\cdot(h_xS_j^x\otimes{\sf I}_{j+1})}{2} \nonumber \\
& -\frac{(1+\delta_{j+1,L})\cdot({\sf I}_j\otimes h_xS_{j+1}^x)}{2}  \nonumber \\
& -\frac{(1+\delta_{j,1})\cdot(h_zS_j^z\otimes{\sf I}_{j+1})}{2} \nonumber \\
& -\frac{(1+\delta_{j+1,L})\cdot({\sf I}_j\otimes h_zS_{j+1}^z)}{2} \text{,}
\end{align}
where $\otimes$ stands for the tensor product, $\delta_{i,j}$ the Kronecker delta, and ${\sf I}_j$ the identity operator at the $j$-th site.

\par Using METTS algorithm, we can successively generate a set of metts $\{\ket{\psi_{\alpha}}\}_{\alpha=1}^N$ for the Hamiltonian $H(0)$ at the initial time. Considering that the occurrence frequency of $\ket{\psi_{\alpha}}\equiv\ket{\psi(i)}$ is asymptotically equal to ${\cal P}(i)/Z$ as $N\to\infty$, the initial Gibbs canonical density matrix~(\ref{eq_rho}) can be expressed as
\begin{align}
\rho=\lim_{N\to\infty}\frac{1}{N}\sum_{\alpha=1}^N\ket{\psi_{\alpha}}\bra{\psi_{\alpha}} \text{.}
\end{align}
The moment generating function~(\ref{eq_G_s}) is therefore given by
\begin{align}
G(s) & =\lim_{N\to\infty}\frac{1}{N}\sum_{\alpha=1}^N{\rm Tr}\left[\ket{\phi_{\alpha}}\bra{\psi_{\alpha}}\right] \nonumber \\
& =\lim_{N\to\infty}\frac{1}{N}\sum_{\alpha=1}^N\braket{\psi_{\alpha}|\phi_{\alpha}} \text{,}
\end{align}
where
\begin{align}
\ket{\phi_{\alpha}}=U^{\dagger}{\rm e}^{sH(\tau)} U{\rm e}^{-sH(0)}\ket{\psi_{\alpha}}
\end{align}
is calculated with the TEBD algorithm. It is noteworthy that this approach is also capable of calculating the work statistics if the system evolves starting from the ground state. Usually, the Density Matrix Renormalization Group (DMRG)~\cite{White_PhysRevLett_1992, White_PhysRevB_1993} is used to find the ground state for quantum many-body systems. In our approach, however, the gound state can be obtained with imaginary time evolution in the low-temperature limit, i.e., calculated from Eq.~(\ref{eq_METTS}) in the limit $\beta\to\infty$, only if the initial cps $\ket{i}$ has the component of the ground state~\footnote{Due to the round-off error in numerical computation, the component of ground state is inevitably introduced during imaginary time evolution.}. It converges exponentially fast, and is also computationally efficient since the ground state is well represented as an MPS. In addition, there is no need to generate an ensemble of states, in contrast to the finite-temperature case.

\begin{figure}
\begin{minipage}[t]{0.8\hsize}
\resizebox{1.0\hsize}{!}{\includegraphics{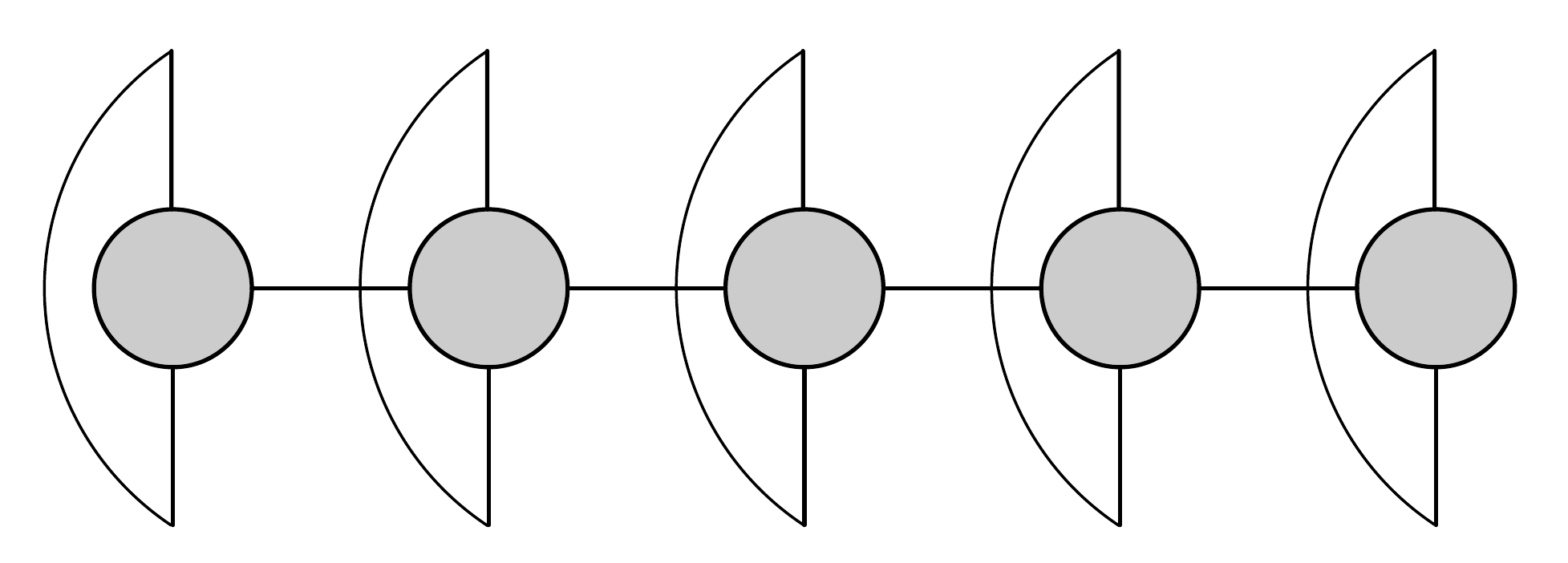}}
\end{minipage}
\caption{The partition function is calculated by tracing out two sets of physical indices of the MPO representing $\exp(-\beta H)$.}
\label{fig_Z}
\end{figure}

\begin{figure}
\begin{minipage}[t]{1.0\hsize}
\resizebox{1.0\hsize}{!}{\includegraphics{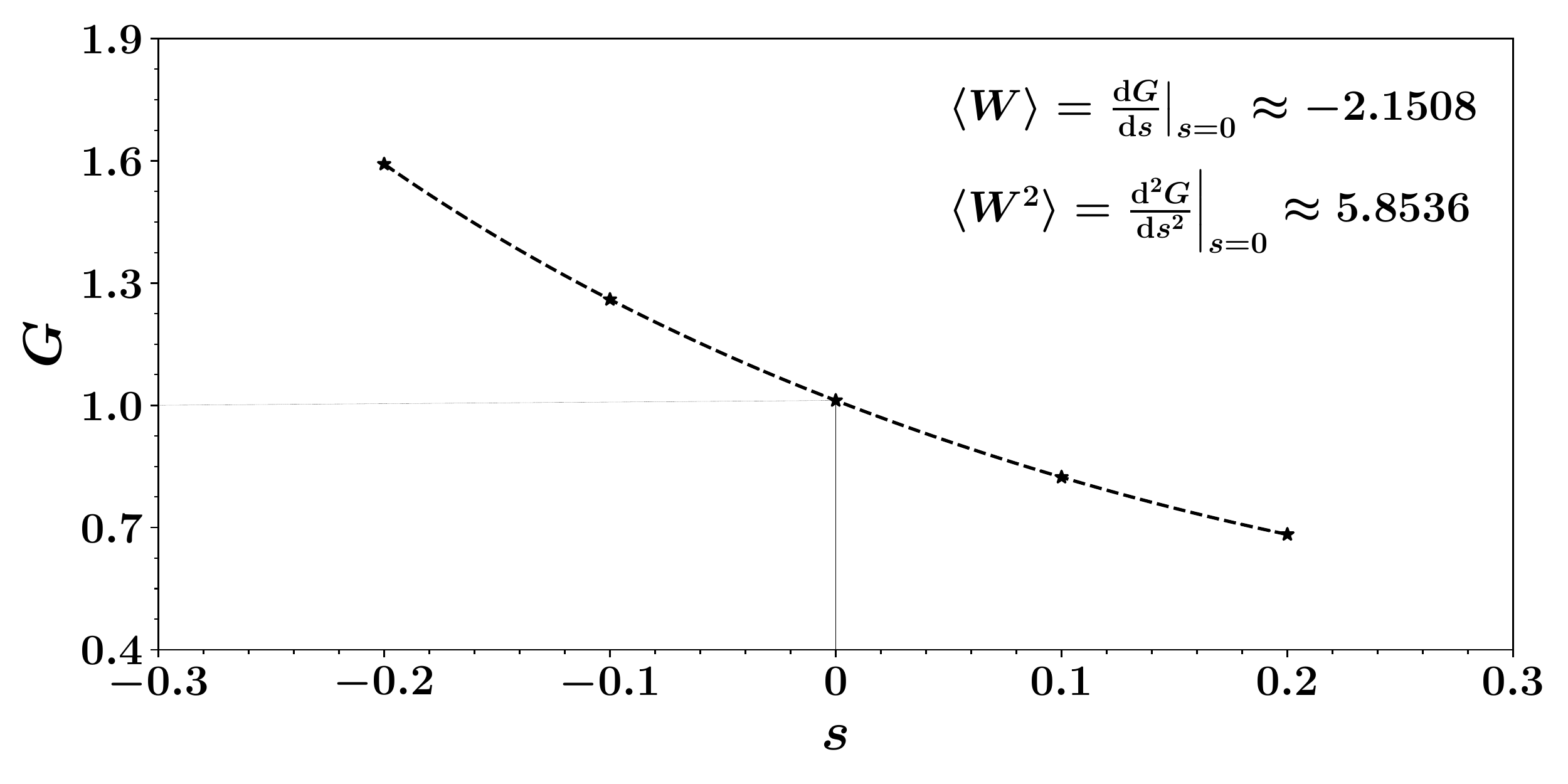}}
\end{minipage}
\caption{Moment generating function $G(s)$. The asterisks are numerical points joined by a dashed line determined from Lagrange interpolation. It looks apparently that $G(0)=1$, as expected. The first and second work moments are evaluated with numerical differentiation. The system is composed of $L=10$ sites, and driven in time from $t=0$ to $t=1$ under the protocol $h_x=t+1$, $h_z=1$. The parameter values $J=\beta=\hbar=1$ are adopted. $10000$ metts are generated in simulation.}
\label{fig_G_s}
\end{figure}

\begin{figure}
\begin{minipage}[t]{1.0\hsize}
\resizebox{1.0\hsize}{!}{\includegraphics{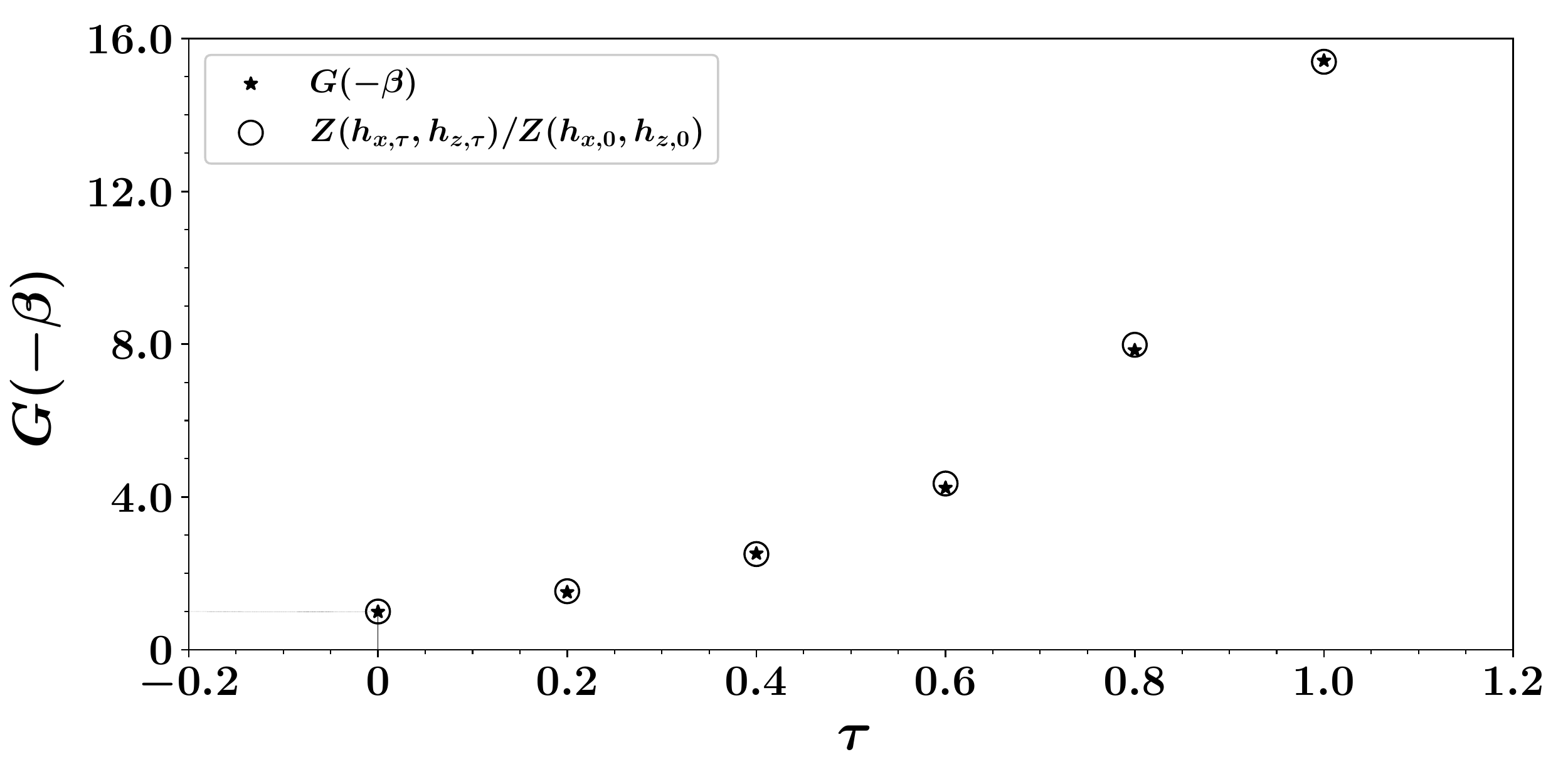}}
\end{minipage}
\caption{Moment generating function $G(-\beta)$ calculated with different time intervals $[0,\,\tau]$, during which the system is driven under the protocol $h_x=t+1$, $h_z=1$. The corresponding values of $Z(h_{x,\tau},h_{z,\tau})/Z(h_{x,0},h_{z,0})$ are also marked.  When there is no driving, i.e., $\tau=0$, then we have $G(-\beta)=1$, as it should be. The system is composed of $L=10$ sites and the parameter values $J=\beta=\hbar=1$ are adopted. $10000$ metts are generated in simulation.}
\label{fig_G_tau}
\end{figure}

\par According to the definition of moment generating function~(\ref{eq_G_s_defined}), we have $\langle\exp(-\beta W)\rangle\equiv G(-\beta)$. To check whether the numerical approach gives the correct work statistics satisfying the Jarzynski equality, the quantity $\exp(-\beta\Delta F)\equiv Z(h_{x,\tau},h_{z,\tau})/Z(h_{x,0},h_{z,0})$ should be calculated in a another way. This can be done through the definition of the partition function, $Z\equiv{\rm Tr}[\exp(-\beta H)]$, where $\exp(-\beta H)$ is expressed as a matrix product operator (MPO). We first prepare an initial identity MPO, $\delta_{j_1,j_1'}\delta_{j_2,j_2'}\cdots\delta_{j_L,j_L'}$, then evolve it under $\exp(-\beta H)$ with the TEBD algorithm to obtain the desired MPO. The tracing operation is now transformed into tensor contraction, see a diagrammatic illustration in Figure~\ref{fig_Z}. Here, it should be pointed out that the Gibbs canonical density matrix can also be constructed directly in this way rather than through generating an ensemble of metts. The advantage of METTS is that there is no need to explicitly calculate the partition function, and also that it can be implemented in parallel computing.

\par Now, we perform numerical simulation. We first calculate $G(s)$ for a fixed driving time interval and several values of argument $s$, see Figure~\ref{fig_G_s}. According to Eq.~(\ref{eq_moments}), the work moments are evaluated with numerical differentiation which can be achieved through Lagrange interpolation~\cite{Press_2007}. We then calculate both $G(-\beta)$ and $Z(h_{x,\tau},h_{z,\tau})/Z(h_{x,0},h_{z,0})$ for various driving time intervals $[0,\,\tau]$ and compare their values. The results are shown in Figure~\ref{fig_G_tau}, from which we see striking agreement. Therefore, the Jarzynski equality, $\langle\exp(-\beta W)\rangle=\exp(-\beta\Delta F)$, is tested, manifesting the reliability of our numerical approach.

\begin{table}
\caption{The numerical results testing the Eq.~(\ref{eq_functional_relation}) with three cases of $\lambda(t)$. $A$ represents the l.h.s., $B$, $C$ respectively ${\rm e}^{-\beta\Delta F}$ and the other part of r.h.s. The system is composed of $L=10$ sites, and is driven in time under the protocol $h_x=t+1$ (respectively, $h_x=1.5-t$), $h_z=1$ from $0$ to $\tau=0.5$ for the forward (respectively, reversed) process. The parameter values $J=\beta=\hbar=1$ are adopted. 5000 metts are generated for each ensemble.}
\begin{center}
\begin{tabular}{>{\centering\arraybackslash}m{2cm}|>{\centering\arraybackslash}m{1.3cm}|>{\centering\arraybackslash}m{1.3cm}|>{\centering\arraybackslash}m{1.3cm}|>{\centering\arraybackslash}m{1.5cm}}
\hline
\hline
   & $A$   &     $B$   &     $C$   &   $BC/A$  \bigstrut \\ \hline
$\lambda=1$   & 15.821   &     3.277   &     4.754   &    0.984 \bigstrut \\ \hline
$\lambda=t+1$   & 4.636   &     3.277   &     1.432   &     1.012 \bigstrut \\ \hline
$\lambda=t^2+t+1$   & 28.737   &     3.277   &     9.053   &     1.032 \bigstrut \\ \hline
\hline
\end{tabular}
\end{center}
\label{tab_functional_relation}
\end{table}

\par The combination of TEBD and METTS can also accomplish more sophisticated tasks. It was proved in Ref.~\cite{Andrieux_PhysRevLett_2008} a universal quantum work relation, reading
\begin{align}
& \left\langle{\rm e}^{\int_0^{\tau}\lambda(t)O_{\rm F}^{\rm H}(t)\,{\rm d}t}{\rm e}^{-\beta H_{\rm F}^{\rm H}(\tau)}{\rm e}^{\beta H(0)}\right\rangle_{\rm F} \nonumber \\
&\quad\quad\quad\quad\quad\quad = {\rm e}^{-\beta\Delta F}\left\langle{\rm e}^{\int_0^{\tau}\lambda(\tau-t)O_{\rm R}^{\rm H}(t)\,{\rm d}t}\right\rangle_{\rm R} \text{,} \label{eq_functional_relation}
\end{align}
which involves an arbitrary function $\lambda(t)$ and an arbitrary time-independent observable $O$. In this relation, $H_{\rm F}^{\rm H}(t)\equiv U^{\dagger}_{\rm F}(t)H(t)U_{\rm F}(t)$, $O_{\rm F}^{\rm H}(t)\equiv U^{\dagger}_{\rm F}(t)OU_{\rm F}(t)$ are respectively the Hamiltonian and the observable in Heisenberg picture for the forward process. $O_{\rm R}^{\rm H}(t)\equiv U^{\dagger}_{\rm R}(t)OU_{\rm R}(t)$ is the observable in Heisenberg picture for the reversed process. The unitary evolution operators for both processes are given by $U_{\rm F}(t)={\cal T}\exp\left[\int_0^t H(t')/({\rm i}\hbar){\rm d}t'\right]$ and $U_{\rm R}(t)={\cal T}\exp\left[\int_0^t H(\tau-t')/({\rm i}\hbar){\rm d}t'\right]$. The symbols $\langle\cdot\rangle_{\rm F}$ and $\langle\cdot\rangle_{\rm R}$ denote the average over the initial canonical ensemble for the forward and the reversed processes, i.e., the density matrices determined by the Hamiltonian $H(0)$ and $H(\tau)$, respectively. It should be noted here that we have neglected the issues relevant to time reversal, for the sake of subsequent numerical convenience in simulation. In other words, all observables are supposed to be even under time reversal, $\Theta O\Theta=O$. When $\lambda(t)=0$, the relation~(\ref{eq_functional_relation}) reduces to the familiar quantum Jarzynski equality, $\langle{\rm e}^{-\beta H_{\rm F}^{\rm H}(\tau)}{\rm e}^{\beta H(0)} \rangle={\rm e}^{-\beta\Delta F}$, where, in the two-point measurement scheme, the factor inside the braket can be interpreted in terms of work performed on the system during the forward process. With the Ising chain previously considered, we now numerically test the relation~(\ref{eq_functional_relation}). The observable is chosen to be the magnetization the $z$ direction, $O=\sum_jS_j^z$, which can be splitted into odd and even parts in a similar manner to the case of the Hamiltonian. The initial canonical ensemble for the forward and reversed processes are generated with METTS. The operators inside the brakets of Eq.~(\ref{eq_functional_relation}) are calculated with TEBD. The numerical results are presented in Table~\ref{tab_functional_relation} with three cases of the function $\lambda(t)$. The last column lists the ratio between two sides of Eq.~(\ref{eq_functional_relation}). The values are almost equal to $1$, in good agreement with the theory.

\par The computer program for numerical simulation is coded in C++~\cite{Stroustrup_2013} with ITensor library~\cite{Fishman_arXiv_2020}. It is assembled in Ref.~\cite{Psarras_arXiv_2021} a comprehensive and up-to-date snapshot of software for tensor computations.

\par \textit{Conclusion}.-- In this Letter, we have introduced an efficient numerical approach for calculating the work statistics of 1D quantum lattice systems. Two tensor-network techniques, TEBD and METTS, are used respectively for time evolution and generating the initial states in thermal equilibrium. This numerical approach enables the detailed investigation of the work statistics under an arbitrary non-perturbative protocol. Therefore, our numerical approach is expected to find further applications in the design of quantum devices operating in nonequilibrium regimes. The combination of TEBD and METTS is versatile, showing the potential of addressing diverse problems in quantum thermodynamics. Extension of the numerical approach to high-dimensional systems will be considered in the future.

\par The authors thank Dr. Jing Chen for initial guidance into the field of tensor networks, and Professor Pierre Gaspard for discussions and encouragement in this research.     Financial support from National Science Foundation of China under the Grants No. 11775001, 11825001, 12147162 is acknowledged.

\end{document}